\documentstyle[12pt,epsf]{article}

\advance \textheight by \topskip
\begin{document}

\begin{center}
\bf
	  AROUND THE KNEE:\\
       SIMULTANEOUS SURFACE AND UNDERGROUND\\
	     MEASUREMENTS AT BAKSAN.

\end{center}

\begin{center}
	A.E.Chudakov, V.B.Petkov\footnote[1]{E-mail: petkov@neutr.novoch.ru},
	  V.Ya.Poddubny, \fbox{A.V.Voevodsky}.
\end{center}
\begin{center}
\it  Institute for Nuclear Reaserch
   Academy of Sciences, Russia
\end{center}
\begin{center}ABSTRACT\end{center}

    We present the experimental data on the knee, as observed in the
 electromagnatic and high energy muon $(E\mu \ge 230$ Gev) components.\\

INTRODUCTION.\\

The break in the shower size spectrum (the so called "knee") at
about $10^6$ particles was observed first by the MSU group more than
40 years ago. But the nature of the knee is the puzzle up to date.
The astrophysical interpretation is the most natural. It is the
steepening of the primary cosmic ray spectrum at energy
$\approx 2\times 10^{15}$ eV. Reasons for such steepening can be found in
acceleration processes as well as in propagation processes. For
example, there is an energy limitation of the acceleration at the
shock front in SNRs. And also, the containment of cosmic rays in galaxy
due to magnetic fields is decreasing  with encreasing particles energy.
These processes are
leading to more heave primary composition after the knee.
The knee also can be a result of the nuclear fragmentation in the
acceleration region. It is leading to the lighter composition
after the knee.
Alternative interpretation is a change in the hadronic interaction
properties.

The situation around the knee is very complicated now. And solving
of the problem requires accurate measurements of the different EAS
component both to verify the main features of the hadronic physics,
and to measure the variations with energy of the primary spectrum
and composition.\\

THE APPARATUS.\\

 The electromagnetic component in our
 experiments is measured by the "Andyrchy" EAS array~\cite{andyr1,andyr2};
 high energy muon ($E\mu \ge 230$ Gev)
 component is measured by the Baksan Underground Scintillation
 Telescope (BUST)~\cite{pst1}. The location of the "Andyrchy" straight above
 the BUST gives us a possibility for simultaneous measurements of
 both components.

     The EAS array "Andyrchy" consists of 37 plastic scintillation
 detectors with an enclosed area of $4.5\cdot 10^4 m^2$.
 Each scintillator has an area of $1 m^2$, with thickness of 5cm and is
 viewed by one PMT. The distance between detectors horizontally
 (projection) is about 40m, the maximum vertical distance (projection)
 is about 150m (Fig.1). The central detector is located above the
 telescope and the corresponding vertical distance is 360 m.
 The altitude above sea level is 2060 m (atmospheric depth $800gr/cm^2$).
 Each detector is equipped by active thermoregulation. The temperature
 is within $1^{\circ}$C at the PMT (and electronics block) and within
 $3^{\circ}$C at the scintillator~\cite{andyr2}.\\
 The electronics inside each detector consists of:\\
a) DC converter (27 V - 2 kV) and voltage divider for PMT high
voltage supply;\\
b) a suppressor of the afterpulses;\\
c) an amplifier of the anode signal of the PMT;\\
d) a logarithmic Charge-to-Time Converter (CTC) of the PMT signals.\\
The output CTC signal has a duration proportional to logarithm of the
charge at 12th dynode~\cite{andyr3}; its leading edge is formed using
an anode signal and is used for timing measurements~\cite{andyr4}.

The energy deposition measurements is performed in natural units - it is
so called relativistic particles. One relativistic particle (r.p.) is most
probable energy deposition from single cosmic ray particle. For our detector
it is 10.5 Mev. The range of the energy deposition measurements is from
0.5 r.p. (the threshold of the CTC) up to more than 1000 r.p. Figure 2
shows a percent error in the energy deposition measurements in the
detector. It is an experimental result, including all apparatus errors
as well as showers fluctuations.

 Trigger formation and all measurements are performed in a registration
 room, which is placed near a center of the array (length of connection
 cables up to 280 m).
 The shower trigger is produced when 4 or more detectors are fired
 ($\Delta t = 3.2$ $\mu s$).
 The trigger rate is $8.8 s^{-1}.$

\setlength{\unitlength}{1mm}

\begin{figure}[t]
\begin{picture}(170,55)

\put(85,17){\line(1,0){70}}
\put(155,20){\line(-2,1){70}}
\put(155,17){\line(0,1){3}}
\multiput(155,20)(-2,0){23}{\line(-1,0){1}}
\multiput(124,35.5)(-5,2.5){7}{\rule{1mm}{1mm}}
\put(110,52){\small the vertical cross section}
\put(107,48){\small of the BUST and EAS array}
\put(108,17){\rule{3mm}{3mm}}
\put(109.5,28){\vector(0,-1){8}}
\put(109.5,33){\vector(0,1){10}}
\put(106,29){\small 360 m}
\put(95,20){\small BUST}
\put(140,15){\vector(1,0){15}}
\put(128,15){\vector(-1,0){18}}
\put(129,14){\small 550 m}

\put(36,45){\small the projection}
\put(32,41){\small to horizontal plane}

\multiput(10,47)(5,0){3}{\rule{1mm}{1mm}}
\multiput(5,42)(5,0){5}{\rule{1mm}{1mm}}
\multiput(0,37)(5,0){7}{\rule{1mm}{1mm}}
\multiput(0,32)(5,0){7}{\rule{1mm}{1mm}}
\multiput(0,27)(5,0){7}{\rule{1mm}{1mm}}
\multiput(5,22)(5,0){5}{\rule{1mm}{1mm}}
\multiput(10,17)(5,0){3}{\rule{1mm}{1mm}}

\put(40,5){\it Fig.1. The EAS array "Andyrchy".}
\end{picture}
\end{figure}

     The BUST is a large underground
 installation with dimensions $16 m \times 16 m \times 11 m$~\cite{pst1}.
 The BUST consists of 3150 liquid scintillation detectors. Each detector
 has dimensions $0.7m \times 0.7 m \times 0.3 m.$
 The coincidence trigger is produced when one or more muons crossing
 the telescope ($\approx 12s^{-1}$) coincide with shower trigger; the
 coincidence rate is $0.1s^{-1}.$
For used selection conditions (near vertical events) the threshold
energy of muons is 230 Gev.\\

 THE EAS SIZE SPECTRUM.\\

In this analysis we used the following EAS selection conditions:\\
1) near vertical events (with zenith angles $\theta \le 10^{\circ} $,
the mean atmospheric depth is 805 $gr/cm^2$).\\
2) only the showers with axis inside of central part of array
(the distance from the central detector not more than 70 $m$)
were included in analysis.\\
EAS arrival directions are obtained from the times of flights among
the different detectors. For used events the angular resolution is
$1.9^{\circ}$.
The reconstruction of shower parameters is performed in units of relativistic
particles. The shower size $N_{r.p.}$, the slope of the lateral distribution
function and the core location are determined by a $\chi^2$  method,
in which the logarithm of the energy deposition in each detector
is compared with the one expected from the NKG lateral distribution function
$$
 \rho(r)=N_{r.p.}{C(s)\over r_0^2} \left({r\over r_0}\right)^{(s-2)}
{\left(1+{r\over r_0}\right)}^{(s-4.5)}
$$
with $r_0 = 96 m$.
The NKG function reproduce with a good accuracy
the experimental data~\cite{andyr5}.
The accuracy of reconstruction is calculated by analysing data obtained
from a simulation that includes the experimental dispersion.
Figure 3 shows a percent error in the size determination versus the size.
 Figure 4 shows the differential size spectrum in r.p. units for
 $7\cdot10^5\le N_{r.p.}\le2\cdot10^7$., where
the shower parameters and size spectrum are reconstructed without
distortions.
The effective running time is $5.02\cdot 10^7$ s (580.9 days).
The steepening of the spectrum is observed at $N_{r.p.}\approx 2\cdot10^6$.

The standart definition of the shower size $N_e$ is the total number of the
charged particles (mainly $e^{\pm}$) at the level of observations.
The measured size $N_{r.p.}$ is the total energy
deposition in infinite detector.
The relations between $N_{r.p}$ and $N_e$ for different primaries
(protons, irons and gammas) was obtained by means of
a simulation of the shower particles crossing the detectors and their housing
for vertical direction. The input data for this simulation are taken from the
CORSIKA 4.50 EAS simulation, the threshold $e^{\pm}$ energy is 5 MeV.
Figure 5 shows the differential size spectra after conversion from
$N_{r.p.}$ to $N_e$ by assumption different primaries (gammas,
protons and irons from top to down).
One can see that the conversion from the total energy deposition
$N_{r.p.}$ to the number of charged
particles $N_e$ is sufficiently ambiguous procedure.
\\

 THE $\overline N{\mu}(N_{r.p.})$ DEPENDENCE.\\
The underground telescope records only a part from the total number of shower
muons, therefore for ${\overline N}_{\mu}$ determination the method
of the mean shower have been used~\cite{msu1,andyr6}.
Events for fixed $N_{r.p.}$
are binned in 10 distance intervals with
$\Delta R=10$ m ($R$ is the distance between  shower's core and
underground telescope). For each interval the number of muons
in the telescope is:
$$M(R_i,N_{r.p}) = \sum_{j=1}^{K}{m_j}$$
(i=1,2,..10), where $K=K(R_i,N_{r.p})$ is the number of showers
and $m_j$ is the number of muons in the telescope for j-th shower.
To explain it more clear - the number of muons in the telescope for
concrete shower can be from 0 up to $\approx$ 150.
Therefore, $${\overline n}(R_i,N_{r.p})=M(R_i,N_{r.p})/K(R_i,N_{r.p})$$ is
the mean number of muons
for showers with size $N_{r.p}$ and distance $R_i$.
The mean number of muons for showers with size $N_{r.p}$ is:
 $${\overline N}_{\mu}(N_{r.p})=
 (1/S_t)\sum_i {\overline n}(R_i,N_{r.p})\cdot S_r(R_i)$$
where  $S_t$ is the effective telescope area (200 $m^2$)
and  $S_r(R_i)$ is ring area.\\
 Fig.6 shows the
 dependence of the mean number of muons in the shower $\overline N\mu$
 on $N_{r.p.}$ for $7\cdot10^5\le N_{r.p.}\le2\cdot10^7$, the effective
 running time is $4.28\cdot 10^7$ s (495.1 days).
One can see some breakdown of single power law
$\overline N\mu \sim {N^{\alpha}}_{r.p.}$
close to the knee ($N_{r.p.}\approx 2\cdot10^6$).\\

ACKNOWLEDGMENT.\\
 The work was supported in part by the Russian Foundation for Basic
Research, project nos. 97-02-17453 and 99-02-16146, and by the Program for
Support of Scientific Schools, grant no. 97-15-96589.
We thank V.V.Alexeenko for helpful comments.

\epsfbox{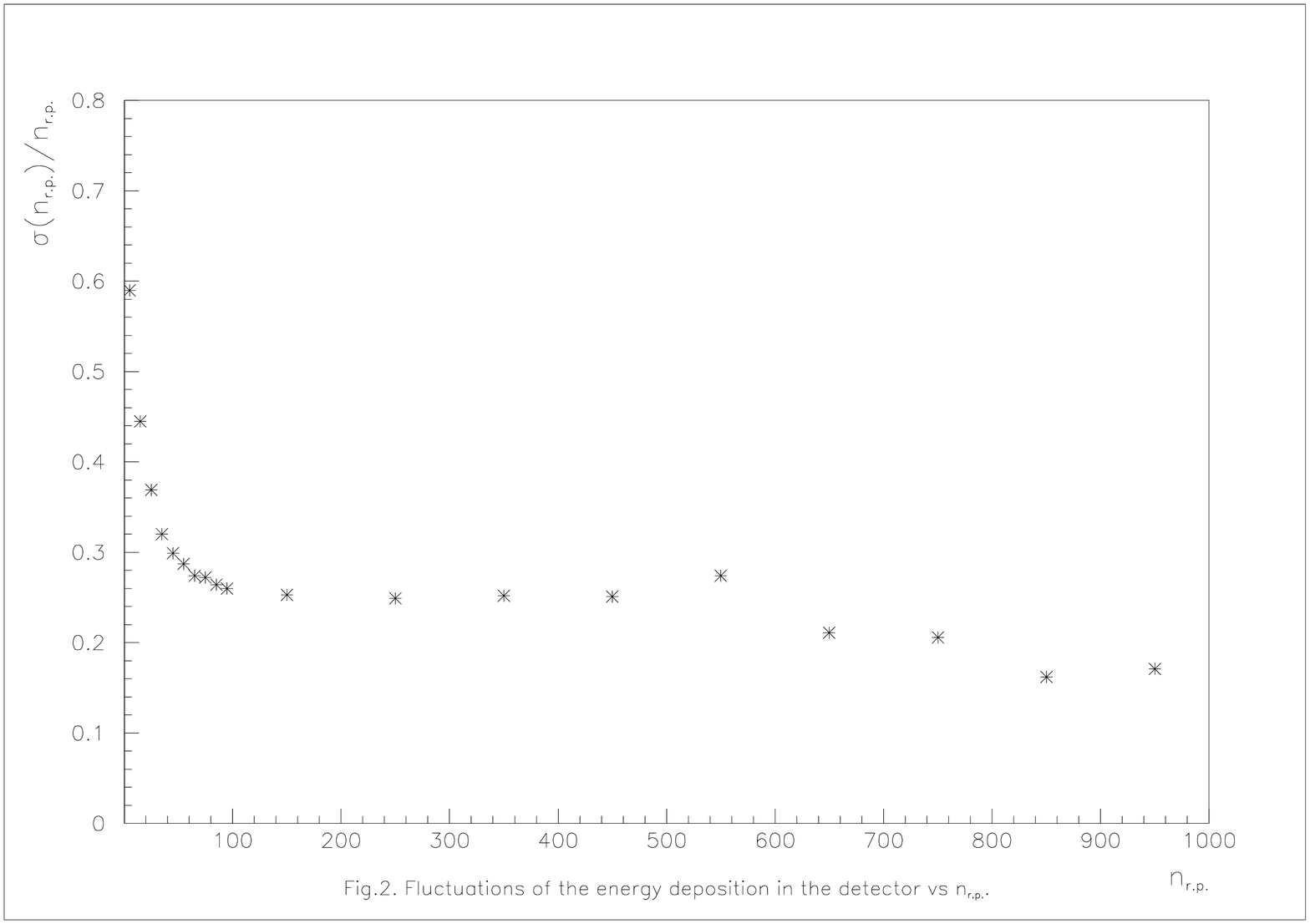}
\epsfbox{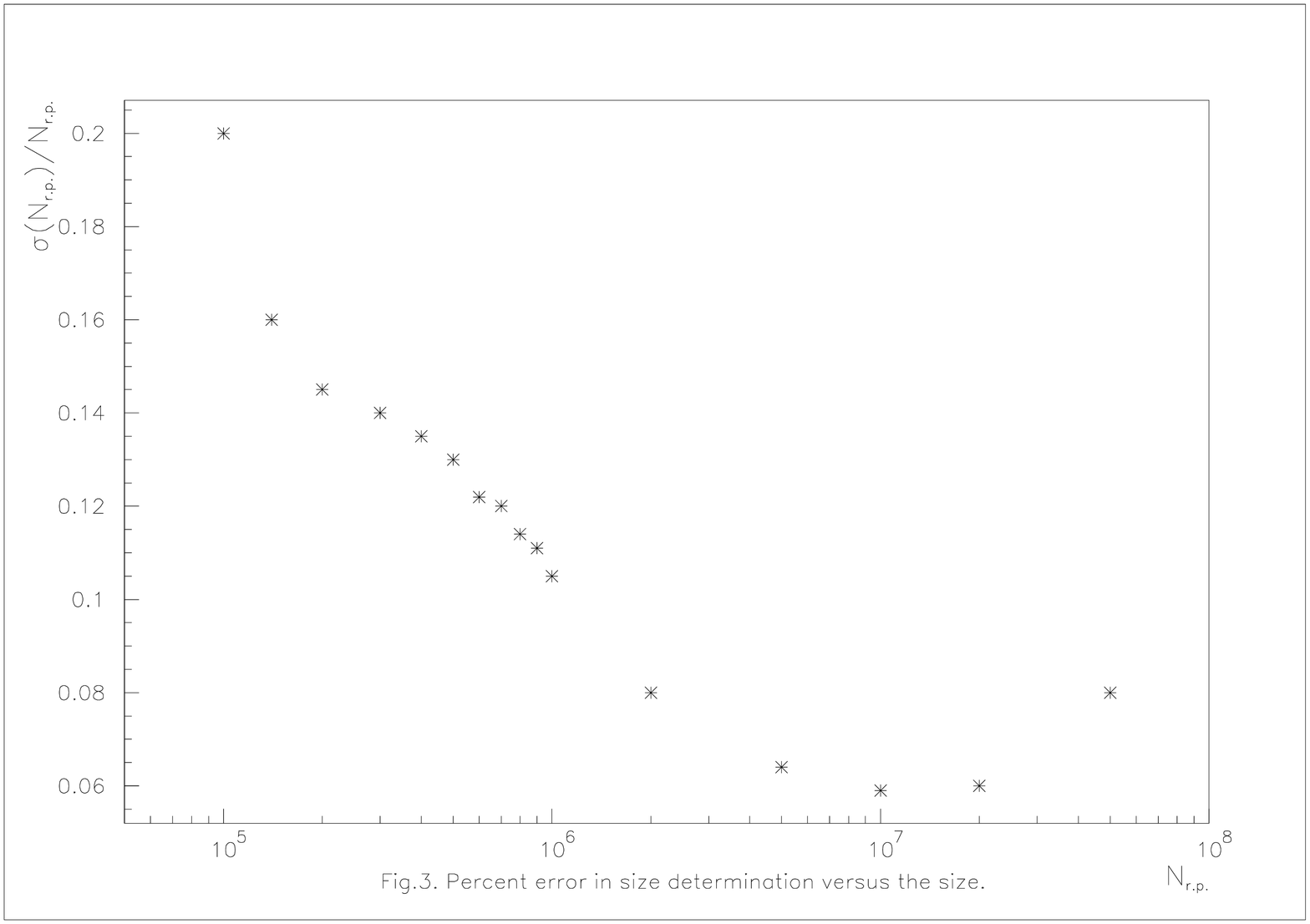}
\epsfbox{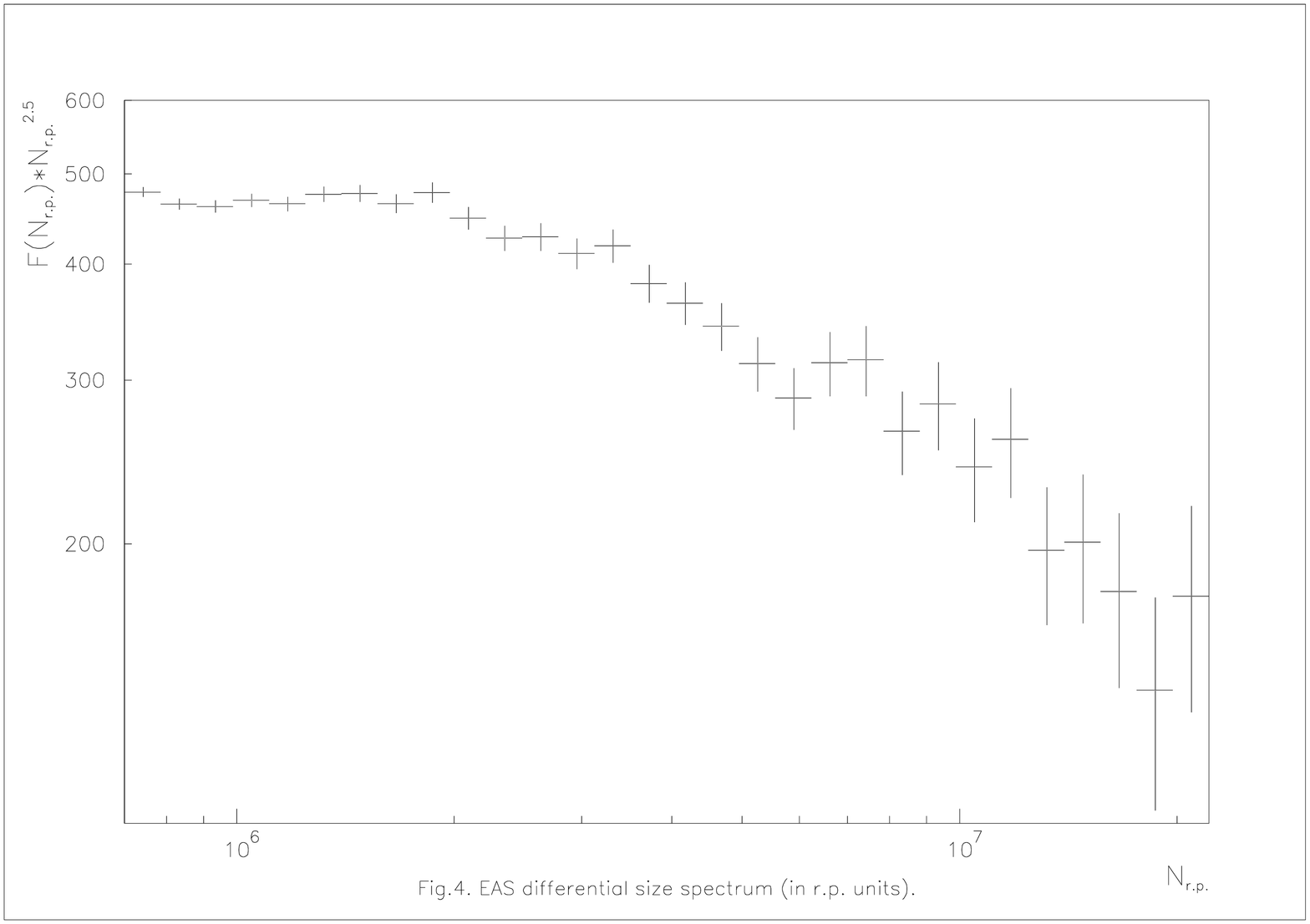}
\epsfbox{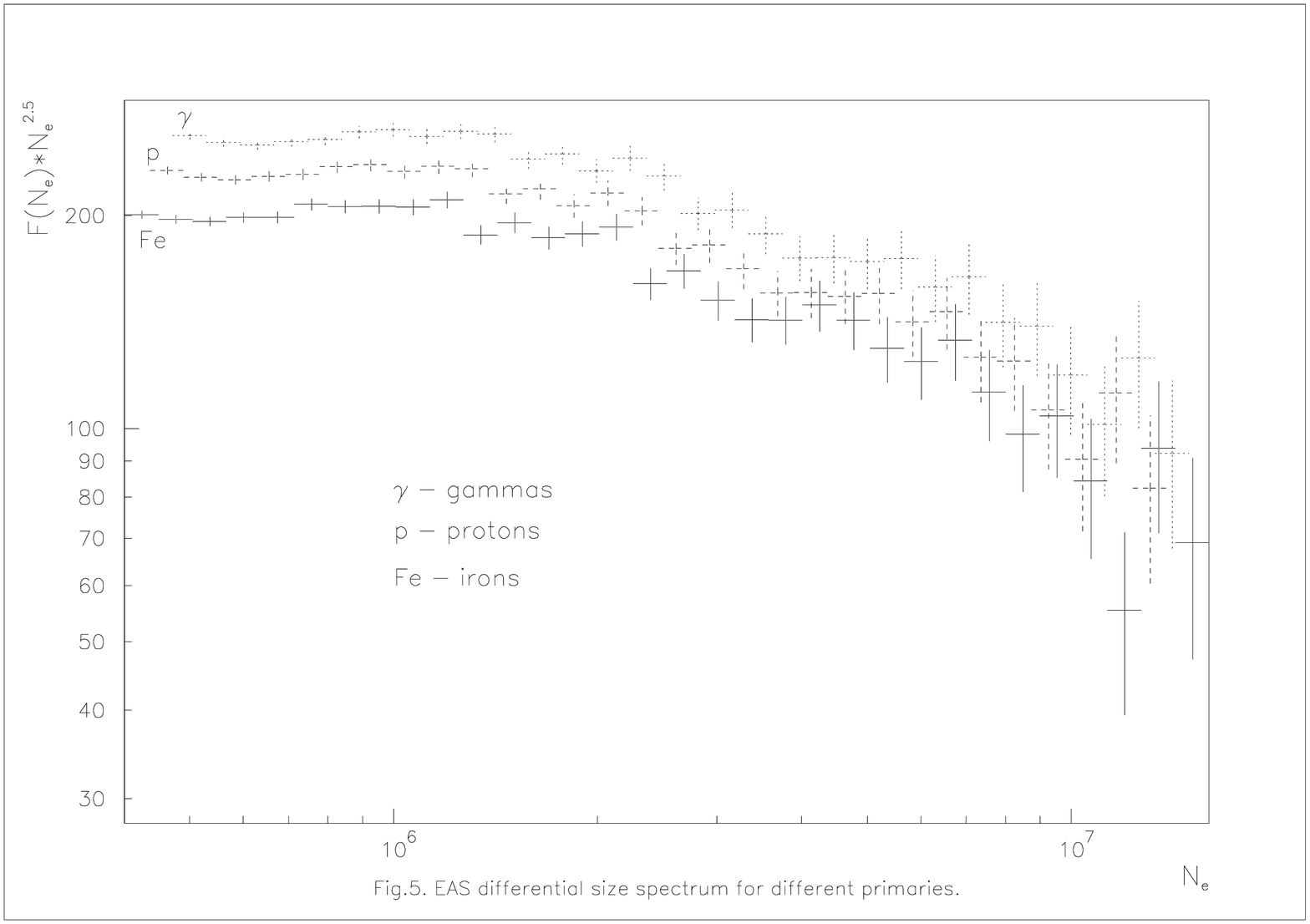}
\epsfbox{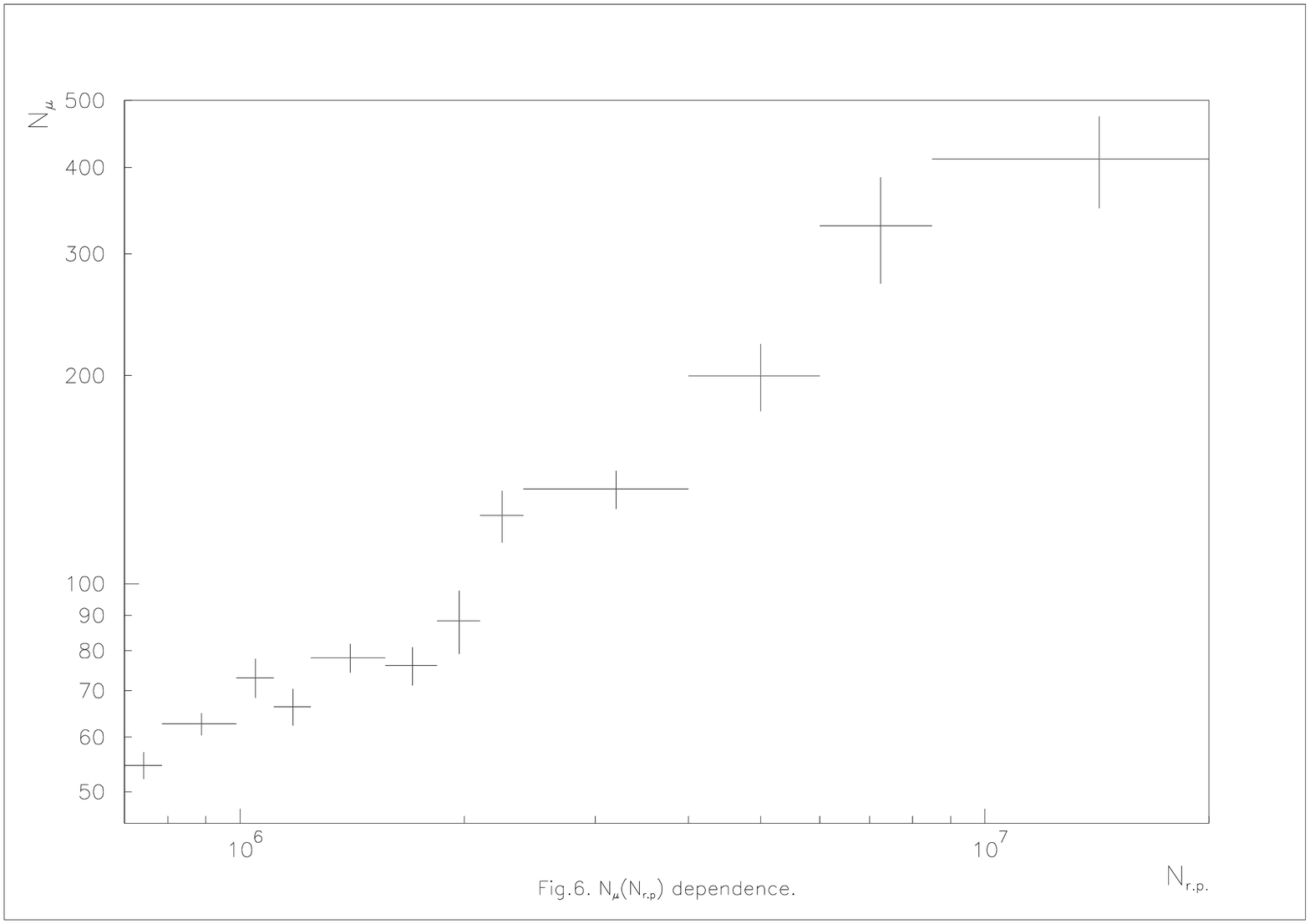}


\begin{thebibliography}{99}
\bibitem{andyr1}
E.N.Alexeyev  et al. {\it Proc. 23rd ICRC}, 2, 474, Calgary (1993)\\
\bibitem{andyr2}
E.N.Alexeyev  et al. {\it Preprint INR	854/94}\\
\bibitem{pst1}
E.N.Alexeyev et al. {\it Proc. 16th ICRC}, 10, 276, Kyoto (1979)\\
\bibitem{andyr3}
V.I.Volchenko  et al. {\it Preprint INR  0913/96}\\
\bibitem{andyr4}
A.V.Voevodsky  et al. {\it Preprint INR  1001/98}\\
\bibitem{andyr5}
A.E.Chudakov et al. {\it Proc. 25th ICRC}, 6, 177, Durban (1997)\\
\bibitem{msu1}
V.V Vashkevich et al. {\it Yad. phys.}, 47, 4, 1054 (1988).\\
\bibitem{andyr6}
A.E.Chudakov et al. {\it Proc. 25th ICRC}, 6, 173, Durban (1997)\\

\end{thebibliography}
\end{document}